\documentclass[useAMS,usenatbib]{mn2e}
\usepackage{bm, amsmath, amssymb, graphicx}
\usepackage[english]{babel}
\title[The orbital periods of HP Lib and V803 Cen]{On the orbital periods of the AM CVn stars HP Librae and V803 Centauri}
\author[G.\,H.\,A. Roelofs et al.]{G.\,H.\,A.~Roelofs,$^1$\thanks{E-mail: g.roelofs@astro.ru.nl} P.\,J.~Groot,$^1$ G.~Nelemans,$^1$ T.\,R.~Marsh$^2$ and D.~Steeghs$^3$\\
$^1$Department of Astrophysics/IMAPP, Radboud University, Toernooiveld 1, 6525 ED Nijmegen, The Netherlands\\
$^2$Department of Physics, University of Warwick, Coventry CV4 7AL, UK\\
$^3$Harvard-Smithsonian Center for Astrophysics, 60 Garden Street, Cambridge, MA 02318, USA}

\newcommand{\obj}{V803 Cen}
\newcommand{\pv}{$P_\mathrm{V803\,Cen}=1596.4\pm1.2\,\mathrm{s}$}
\newcommand{\ph}{$P_\mathrm{HP\,Lib}=1102.8\pm0.2\,\mathrm{s}$}
\begin{document}
\maketitle

\begin{abstract}
We analyse high-time-resolution spectroscopy of the AM CVn stars HP Librae and V803 Centauri, taken with the New Technology Telescope (NTT) and the Very Large Telescope (VLT) of the European Southern Observatory, Chile.

We present evidence that the literature value for V803 Cen's orbital period is incorrect, based on an observed `S-wave' in the binary's spectrogram. We measure a spectroscopic period \pv\ of the S-wave feature, which is significantly shorter than the 1611-second periods found in previous photometric studies. We conclude that the latter period likely represents a `superhump'. If one assumes that our S-wave period is the orbital period, V803 Cen's mass ratio can be expected to be much less extreme than previously thought, at $q\sim 0.07$ rather than $q\sim 0.016$. This relaxes the constraints on the masses of the components considerably: the donor star does then not need to be fully degenerate, and the mass of the accreting white dwarf no longer has to be very close to the Chandrasekhar limit.

For HP Lib, we similarly measure a spectroscopic period \ph. This supports the identification of HP Lib's photometric periods found in the literature, and the constraints upon the masses derived from them.

\end{abstract}


\section{Introduction}

V803 Centauri and HP Librae are both members of the class of AM CVn stars: ultra-compact binaries with orbital periods below about one hour, in which a white dwarf accretes helium-rich matter from a degenerate or semi-degenerate companion. They are of interest for binary evolution theory, being a distinct class of evolutionary end-products that have probably undergone a common-envelope phase twice. Their ultra-short orbital periods make them important sources for gravitational-wave astronomy; they are the strongest (and so far only) known sources for future detectors such as the \emph{Laser Interferometer Space Antenna (LISA)} (e.g.\ \citealt{npy,roelofsamcvn}).

However, determining their orbital periods is made difficult by the superhump phenomenon, by which the observed photometric periods do not match with their orbital periods, which is thought to be due to tidal interactions between the accretion disc and the donor star \citep{whitehurst88}. Because of this complication, several of the short-period AM CVn stars do not have a secure orbital period measurement. AM CVn itself, which has a number of photometric periods dominated by one at 1051 seconds, has been shown from spectroscopy to have an underlying 1029-second orbital period \citep{nsg}. This is based on an observed `S-wave' feature that is commonly thought to be associated with the region of impact of the accretion stream into the accretion disc in interacting binaries. Similar behaviour may be expected for HP Lib, a system virtually identical to AM CVn in appearance, and for V803 Cen, which appears similar to AM CVn and HP Lib most of the time but shows occasional drops in brightness of up to five magnitudes, presumably caused by a thermal instability in its accretion disc \citep{osaki}.

V803 Cen was discovered as a strongly variable and very blue source by \citet{elvius}, in a photometric survey of the region around the galaxy NGC\,5128. \citet{donoghue87} obtained high-speed photometry which showed the source to be a rapid variable, exhibiting variations on a $\sim$1611-second period. They noted the similarities with another recently discovered rapid variable, CR Boo, and classified it as an interacting binary white dwarf; the 1611-second period was recovered several times in follow-up photometry \citep{donoghue89,donoghue90}. An extensive monitoring campaign by \citet{patterson00} uncovered a 1618-second signal whenever V803 Cen was in a state of high brightness, while recovering the 1611-second variations in low and intermediate brightness states. Although \citet{patterson00} initially interpreted these results in terms of a slowly dying superhump that persists into the low state, the suggestion that 1611\,s is in fact the orbital period appeared shortly thereafter \citep{patterson01}. This allowed for an estimate of the mass ratio $q=M_2/M_1$ of the binary based on the observed relation between the superhump period excess, $\epsilon\equiv \left(P_\mathrm{sh}-P_\mathrm{orb}\right)/P_\mathrm{orb}$, and the mass ratio in Cataclysmic Variables (e.g.\ \citealt{patterson05}).

V803 Cen's tiny superhump period excess then implies an extremely small mass ratio $q\approx0.016$. As noted by \citet{deloye}, the minimum mass for the donor star, which corresponds to the mass of a cold, fully degenerate white dwarf that fills its Roche lobe in V803 Cen's orbit, implies that the accretor must be massive: $M_\mathrm{1,V803\,Cen}\gtrsim 1.3\,M_\odot$. Conversely, the Chandrasekhar limit to the mass of the accretor then indicates that the donor in V803 Cen has to be nearly fully degenerate, which would suggest that V803 Cen was formed from a double-degenerate rather than from a single-degenerate binary (see \citealt{nelemans}).

In order to explain the large observed luminosity of V803 Cen \citep{roelofshst}, while requiring that the donor star be fully degenerate, one would require an accretor mass even closer to the Chandrasekhar limit in order to extract a sufficient amount of energy per unit mass of accreted material: $M_\mathrm{1,V803\,Cen}\approx1.43M_\odot$, if one assumes the mass transfer rate to be set by the system's gravitational-wave losses. Given these spectacular implications, a more secure identification of the orbital period of the binary is desirable. In this paper we search for a spectroscopic signal of V803 Cen's orbital period in observations made with the Very Large Telescope (VLT), Chile.

The second object discussed in this paper, HP Lib, was discovered more recently in the Edinburgh-Cape survey for blue objects and recognised immediately as a twin of the prototype star AM CVn. It exhibits 1119-second brightness variations, but no large outbursts like the ones observed in V803 Cen and CR Boo \citep{donoghue94}.

\citet{patterson02} again conducted an extensive campaign of time-series photometry and discovered, in addition to the main 1119-second variations, an extremely low-amplitude modulation on $1102.70\pm0.05$ seconds. This was readily interpreted as the likely orbital period of the binary, while the resulting constraints on the masses of the components, inferred from the (larger) superhump period excess \citep{deloye}, are not nearly as stringent as those derived for V803 Cen. We here analyse fast spectroscopy taken with the New Technology Telescope (NTT) in order to, again, search for a spectroscopic signal of the orbital period of the binary.

\section{Observations and data reduction}

\begin{table*}
\begin{center}
\begin{tabular}{l l l r r r r r}
\hline
Night		&UT		        &Set-up                 &Spectral range         &Resolution     &Exposures      &Exp.\ time  &Typical seeing\\
                &                       &                       &(\AA)                  &(\AA)          &               &(s)            &('')\\
\hline
\hline
V803 Cen        &&&&&\\
2005/03/01	&05:20--06:15	        &VLT+FORS2/1400V        &4625--5930             &$\sim$1.8       &50             &30             &0.6\\
        	&07:35--08:30	        &VLT+FORS2/1200R        &5870--7360             &''              &50             &30             &0.6\\
2005/03/02	&05:43--09:43	        &VLT+FORS2/1200R        &''                     &''              &200            &30             &0.6\\
2004/03/09      &02:00--04:20           &NTT+EMMI/\#6           &4330--5050             &$\sim$0.7       &10             &750            &0.7\\        
2004/03/10      &02:14--02:44           &NTT+EMMI/\#6           &''                     &''              &2              &900            &0.7\\        
\hline
HP Lib          &&&&&\\
2004/03/06      &05:09--10:04           &NTT+EMMI/\#6           &''                     &''              &194            &55             &1.0\\        
2004/03/07      &04:46--10:03           &NTT+EMMI/\#6           &''                     &''              &191            &55             &0.9\\        
2004/03/08      &04:50--10:02           &NTT+EMMI/\#6           &''                     &''              &223            &45--55         &0.7\\        
2004/03/09      &04:55--10:00           &NTT+EMMI/\#6           &''                     &''              &225            &45             &0.7\\        
2004/03/10      &04:49--10:05           &NTT+EMMI/\#6           &''                     &''              &237            &45--50         &0.7\\        
\hline
\end{tabular}
\caption{Summary of our observations of V803 Cen and HP Lib. The set-up indicates the telescope, the instrument and the grism or grating.}
\label{observations}
\end{center}
\end{table*}

We obtained phase-resolved spectroscopy of \obj\ on 1 and 2 March 2005 with the Very Large Telescope (VLT) and the FOcal Reducer/low dispersion Spectrograph (FORS2). The observations consist of 50 spectra taken with the 1400V grism, covering 4625--5930\,\AA, followed by 250 spectra with the 1200R grism, covering 5870--7360\,\AA, all with a 30-second exposure time. Weather conditions were good, with a median seeing around $0.6''$, giving an effective resolution of about 1.8\,\AA\ or 80 km/s at 6678\,\AA.

All spectra were taken with a 1$''$ slit. The detector was the standard MIT mosaic consisting of two CCDs of 2k$\times$4k, 15$\mu$m pixels, binned by a factor of two. The dispersion solution was obtained from a set of 42 night sky lines for the 1200R grism, and from standard HeNeAr arc exposures for the 1400V grism, resulting in $\sim$0.1\,\AA\ root-mean-square residuals. In order to take full advantage of the good seeing, we corrected the 1200R spectra for the possible drifting of the star within the slit and the consequent drift in the dispersion solution (to zeroth order) using the sharp atmospheric absorption edge at 6868\,\AA. All spectra were transformed to the heliocentric rest-frame prior to analysis; in as far as possible we have verified that the position and time stamps on the data were correct. The average spectrum was corrected for instrumental response using spectrophotometric standard star EG274.

In addition to the phase-resolved spectra, we obtained spectra of long exposures with the New Technology Telescope (NTT) and the ESO Multi-Mode Instrument (EMMI) on 9 and 10 March 2004. We used grating \#6 and a 1$''$ slit to obtain an effective resolution of 0.7\,\AA\ FWHM over the range 4330--5050\,\AA. The detector was the MIT mosaic, consisting of two CCDs of 2k$\times$4k, 0.15$\mu$m pixels, binned by a factor of two.

The same NTT set-up was used to obtain phase-resolved spectra of our second target, HP Lib, on five consecutive nights starting March 6, 2004. The exposure times were 45--55 seconds and the weather conditions were again good, with a typical seeing of 1.0$''$ on the first night, dropping to 0.7$''$ on the subsequent nights. The wavelength calibration was done with HeAr arc exposures, in which a total of about 28 arc lines were fitted with a fourth-order Legendre polynomial, leaving 0.02\,\AA\ residuals. All spectra were again transformed to the heliocentric rest-frame. The average spectra were flux-calibrated with the standard star HD49798.

A summary of our observations is given in table \ref{observations}.

\section{Results}

\subsection{Average and phase-binned spectra}
\label{averagespectrum}

\begin{figure*}
\centering
\includegraphics[angle=270,width=\textwidth]{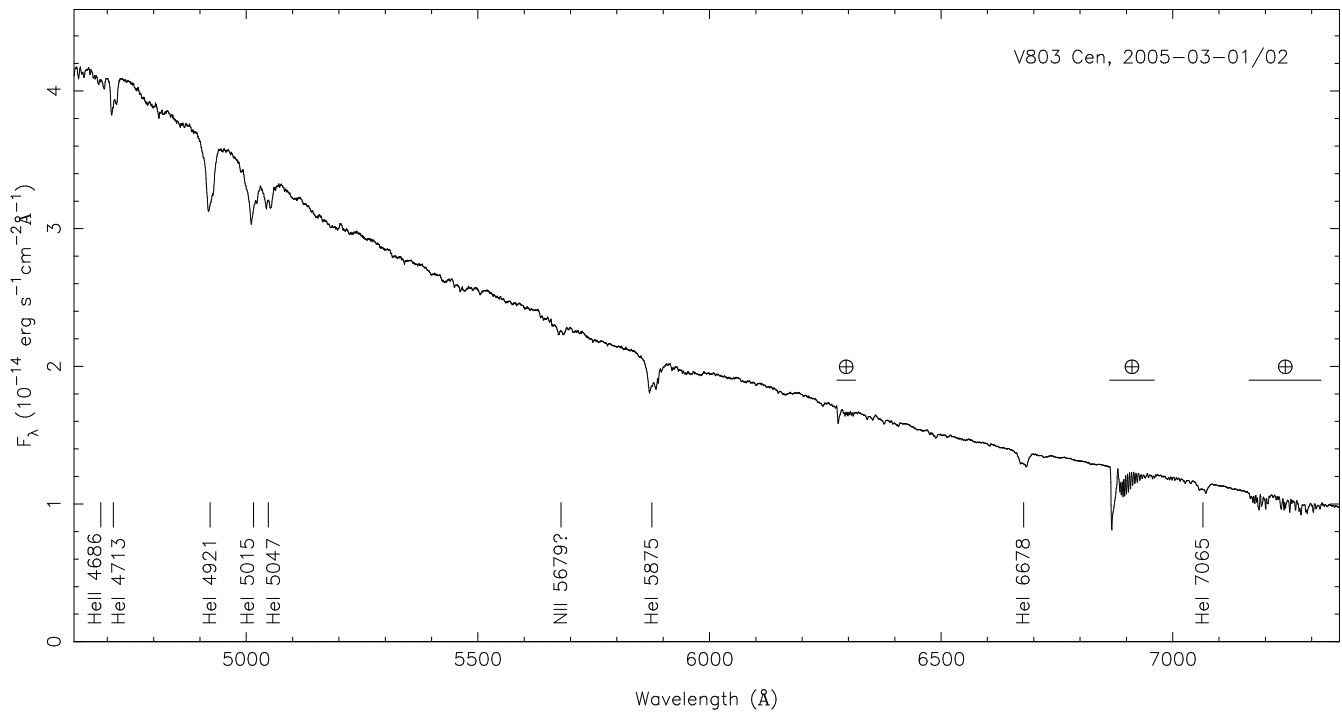}
\caption{Average visual and red spectrum of V803 Cen observed with the VLT, with labels attached to the strongest features. The spectrum of the second night has been scaled up by $\sim$5\% to match the intensity of the first night's spectrum.}
\label{averageV803CenVLT}
\end{figure*}

We caught V803 Cen in a high state on the VLT, where we measured an average magnitude of about $V=13.0$ from our spectroscopy (see \citet{patterson00} for a description of V803 Cen's different brightness states). Its average spectrum shows mainly neutral helium lines in absorption (see figure \ref{averageV803CenVLT}), but in a trailed spectrum we find weak additional lines of \mbox{Si\,{\sc ii}} 6346 and \mbox{Si\,{\sc ii}} 6372, also seen in other AM CVn stars \citep{groot,roelofs}, and two features which coincide with \mbox{N\,{\sc ii}} 5679 and \mbox{N\,{\sc ii}} 6482. There is no evidence for any hydrogen in the spectrum of V803 Cen.

\begin{figure}
\centering
\includegraphics[angle=270,width=84mm]{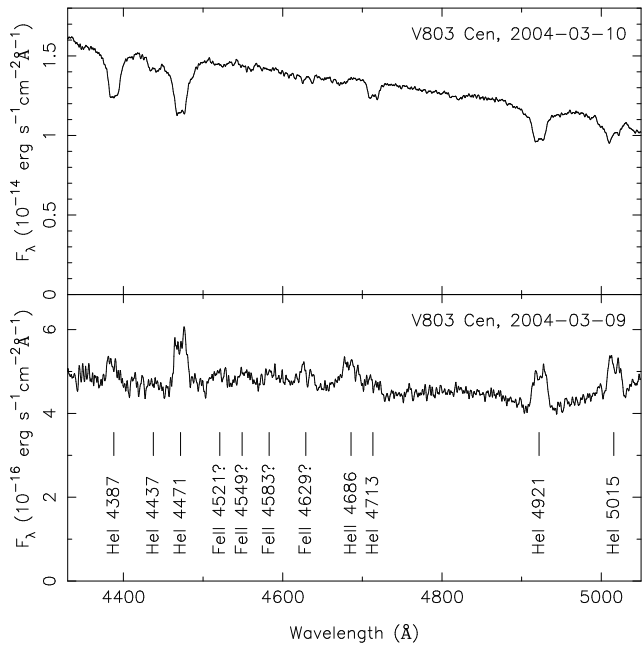}
\caption{Average blue spectra of V803 Cen observed with the NTT.}
\label{averageV803CenNTT}
\end{figure}

On the NTT, V803 Cen was observed to be in a low state, close to $V=17.2$, on March 9, 2004. The average spectrum shows double-peaked emission lines, characteristic of an accretion disc. The strongest lines are due to neutral helium, but \mbox{He\,{\sc ii}} 4686 is also clearly present and there appears to be a group of lines at 4521\,\AA, 4549\,\AA, 4583\,\AA\ and 4629\,\AA\ that may represent iron (see figure \ref{averageV803CenNTT}). The next night, on March 10, the star had increased in brightness by about three magnitudes, to $V=14$, and changed from an emission-line to an absorption-line spectrum, as also shown in figure \ref{averageV803CenNTT}.

\begin{figure}
\centering
\includegraphics[angle=270,width=84mm]{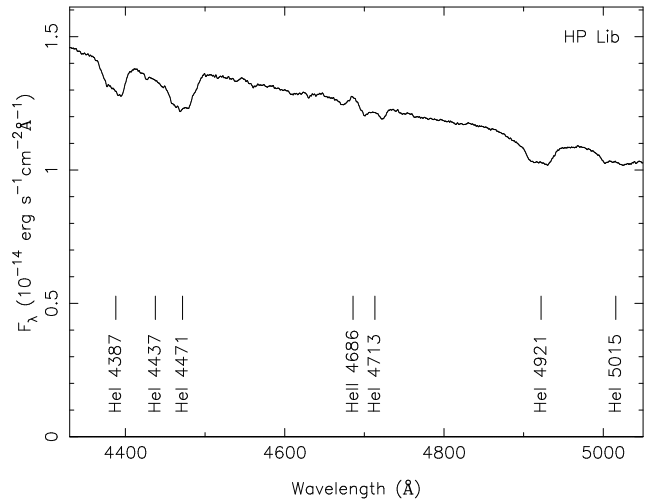}
\caption{Average blue spectrum of HP Lib observed with the NTT.}
\label{averageHPLib}
\end{figure}

HP Lib has an absorption-line spectrum similar to that of V803 Cen in the high state, showing mainly neutral helium lines as well as weak emission of \mbox{He\,{\sc ii}} 4686 (on top of apparent broad absorption of the same line). See figure \ref{averageHPLib}. All in all, the average spectra of HP Lib and of V803 Cen in the high state are very similar to that of AM CVn \citep{roelofsamcvn}.

\subsection{Doppler tomography}
\label{tomography}

Doppler tomography \citep{dopplermapping} is a powerful technique for measuring the orbital period of interacting binaries: especially the ones that show no periodicity in broad-band photometry, and the ones that show a broad-band signal dominated by superhumps (e.g.\ \citealt{nsg}). Assuming only that there is an emission region that stands out and is stationary in the binary frame, usually in the form of a `bright spot' caused by the accretion stream--accretion disc impact point, one can track the orbital phase of the binary by tracking the phase of the emission region. In a Doppler map, the phase of the feature (i.e.\ the angle subtended by the feature's $K_X$ and $K_Y$ velocities in polar coordinates, with arbitrary zero-point) will remain the same for a series of data-sets taken at different times, if and only if one phase-folds the data-sets on the true orbital period of the binary. If the phase-folding period is slightly off from the true orbital period, the feature will still show up in individual Doppler maps made from data-sets with a sufficiently short timebase, but slightly smeared out and its phase in the maps will be seen to drift slowly over time.

Here we apply this technique to V803 Cen and HP Lib. When we phase-fold the spectra of V803 Cen from the first or second night on the 1611-second period suggested from photometry, there is a persistent spot-like emission region in the Doppler tomogram. See figure \ref{doppmapV803Cen}. If this represents a real, stationary feature in the binary, it would suggest that 1611\,s is close to the real orbital period of the binary -- if it were far off, the feature would be (too) smeared out when phase-folded in a Doppler tomogram.

However, the phase of this feature is seen to drift by about 180 degrees between the two nights of our observations, indicating that 1611 seconds is not exactly the orbital period \emph{if} the feature is fixed in the binary frame (see figure \ref{doppmapV803Cen}).

We now assume for a moment that the emission spot seen on each of the nights is the same feature and is stationary in the binary frame over this period of time; we shall refer to it as the `bright spot'. We can easily keep track of the bright spot's phase over the four-hour observing baseline that we have on the second night of our observations, starting from the approximate orbital period suggested from photometry, and then extrapolating to the first night without losing count of the orbital cycles. We then adjust the folding period so as to align the bright spot in the Doppler maps of both nights. This yields a spectroscopic period of 1596.4 seconds, which is also the orbital period if we assume that the bright spot is fixed in the binary frame.

We next perform a simple Monte Carlo simulation, using the bootstrap method as described in \citet{roelofsaw} and \citet{roelofsamcvn}, in order to establish the bright spot phase resolution given by our data (at the same time, it allows us to verify that the bright spot we see is in fact a real feature rather than a coincidence of noise). We make a large ensemble of randomised Doppler tomograms from the data that we have on each night, and determine the scatter in the measured bright spot phase. We conclude that we can measure the relative bright spot phase on night 1 and night 2 to an accuracy of 5 degrees, not taking into account any intrinsic drift in the bright spot phase. If we assume that the bright spot corresponds to the conventional accretion stream--accretion disc impact point, and we allow the effective accretion disc radius to change by $0.1 R_{L_1}$ between the two nights (where $R_{L_1}$ is the distance from the centre of the accreting star to the inner Lagrange point), this adds a possible intrinsic drift of the bright spot phase in the binary frame of $\sim$10 degrees. The total error in the phasing of the bright spots is thus no more than 15 degrees. This leads to a spectroscopic period \pv. Figure \ref{doppmapV803Cen} shows the linear back-projection Doppler tomograms that we get when we phase-fold the data of each night on that period. From our Monte Carlo sample, the bright spot velocity semi-amplitude in V803 Cen is measured to be $K_\mathrm{V803\,Cen}=170\pm15$ km/s.

\begin{figure}
\centering
\includegraphics[angle=270,width=84mm]{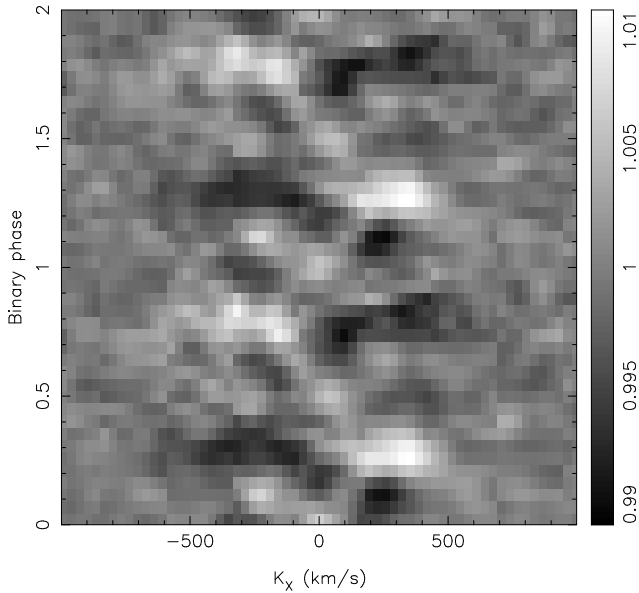}
\caption{Trailed spectrum of the \mbox{He\,{\sc i}} 4921, \mbox{He\,{\sc i}} 5875, \mbox{He\,{\sc i}} 6678 and \mbox{He\,{\sc i}} 7065 lines in V803 Cen, with the average spectrum subtracted and folded on $P_\mathrm{V803\,Cen}=1596.4$\,s. The grey-scale indicates the flux levels relative to the mean spectrum; brighter pixels contain more flux.}
\label{trailV803Cen}
\end{figure}

\begin{figure}
\centering
\includegraphics[angle=270,width=84mm]{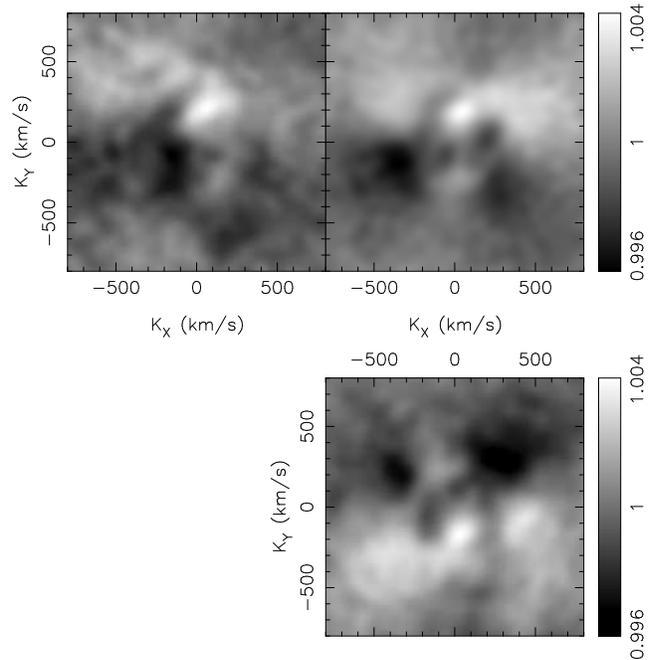}
\caption{Linear back-projection Doppler tomogram corresponding to the trailed spectrum shown in figure \ref{trailV803Cen}, only now split into night 1 (left) and night 2 (right) to show the alignment of the bright spot. As a comparison, the lower right panel shows the shift of the bright spot between the first and second night, if we fold the data on $P=1611$\,s.}
\label{doppmapV803Cen}
\end{figure}

Figure \ref{trailV803Cen} shows V803 Cen's spectrogram that we obtain when we fold the data on the period obtained above. For this spectrogram we used the two hours of spectroscopy that we have on the first night, and combined it with the first and fourth hour of data from the second night in order to obtain a data-subset that is more or less evenly distributed in time. Apart from an `S-wave' that coincides in phase and amplitude with the bright spot seen in the Doppler tomograms of figure \ref{doppmapV803Cen}, there is significant variability in the spectral lines on a period roughly three times shorter than the S-wave period. In addition, the pattern of this variability appears to shift with time relative to the S-wave. We therefore speculate that it may be related to the superhump phenomenon. By using a data-subset that is somewhat evenly distributed in time, as described above, we find that we can minimise the amplitude of this rapid variability and maximise the visibility of the S-wave. Based on the observed changes in the pattern we speculate that it may disappear almost completely if one were to observe V803 Cen for a long time and phase-fold all the spectra (for instance if one were to fully sample the beat period between the orbital and superhump periods); this can however not be verified with our current data-set.

\begin{figure}
\centering
\includegraphics[angle=270,width=84mm]{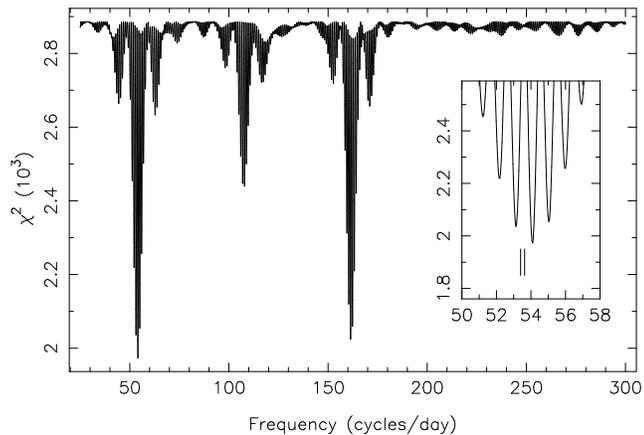}
\caption{Chi-square values of sinusoid fits to normalised red wing/blue wing flux ratios of the helium lines in V803 Cen. The two vertical lines in the close-up indicate the 1618-second (left) and 1611-second periods known from photometry.}
\label{pgramV803Cen}
\end{figure}

As a further check on V803 Cen's spectroscopic period, we compute a periodogram of the flux ratios of the red and blue wings of the spectral lines, as described in \citet{roelofs} following \citet{nather}. We let the wings of the lines extend to $\pm$500 km/s based on the trailed spectrum of figure \ref{trailV803Cen}; for each spectrum we add up the fluxes in the red wings and in the blue wings of all the helium lines. We normalise the red/blue flux ratios such that the average ratios for the spectra taken with the different grisms and on different nights are the same; this removes the potential variability on timescales of a few hours and longer, but also weeds out the systematic differences between spectra taken with different set-ups and on different nights. Note that this also removes the influence of intrinsic, long-period skewness variations in the spectral lines, which have been seen in V803 Cen \citep{donoghue89} as well as in other AM CVn stars to occur on the beat period between the orbital and superhump periods (see \citealt{patterson93}). We model the flux ratios as a function of time with a constant plus a sinusoid, and compute chi-square values for the best-fitting model at each trial frequency. The resulting periodogram is shown in figure \ref{pgramV803Cen}. Although, unsurprisingly, the data are poorly fitted with a constant plus a single sinusoid, shown by the large minimum reduced chi-square value of 6.7, the best-fitting sinusoidal period is $P=1597.6$\,s, in perfect agreement with the period derived from the Doppler tomograms. The nearest aliases occur at $P=1626.3$\,s ($\Delta \chi^2=63$) and $P=1569.8$\,s ($\Delta \chi^2=81$).

The large $\Delta \chi^2$ values indicate that the best-fitting period fits significantly better than the next-best periods. In principle, however, these values can be inflated by underestimated errors on the data points. If we assume for a moment that our simple model should describe the data perfectly, and we renormalise the errors on the data points so as to yield a best-fitting reduced chi-square of 1, the nearest alias would have $\Delta \chi^2=9.5$. The best-fitting period of $P=1597.6$\,s would then still be preferred over the next-best period at the 95\% confidence level (for a chi-square distribution with four degrees of freedom; see \citealt{avni}). This is the worst-case scenario; $\Delta \chi^2$ for the next-best period is most likely much closer to the measured value of 63 since the estimates of our errors (which are shot-noise dominated and should be fairly well-behaved) are probably not that far off.

Lastly, we do a slightly more advanced fit of sinusoids in which we allow second and third harmonics of the test frequencies, with their amplitudes as additional free parameters, to try to include the strong harmonics seen in the periodogram of figure \ref{pgramV803Cen} in the fit. We recover the same best-fitting periods as before, to better than half a second, while the reduced chi-square of the best fit improves from 6.7 to 4.3. The strongest second and third harmonics stay at $P=1602/2$ and $1616/2$ and $P=1606/3$ seconds, suggestive of a superhump origin if one assumes the base period of $P=1597.6$\,s to be orbital. The $\Delta \chi^2$ between the best-fitting and second-best-fitting frequencies improves considerably, from $\Delta \chi^2=63$ to $\Delta \chi^2=537$.

We conclude that our periodograms of sinusoid fits support the period derived from the bright spot phasing in the Doppler tomograms, although we take note of the fact that the baseline of our observations is relatively short (only 28 hours; see table \ref{observations}), so that it will be difficult to claim the same frequency resolution in our periodograms as we have from the bright spot phasing method.

The spectrogram of HP Lib, when folded on the suggested orbital period of $P_\mathrm{HP\,Lib}=1102.7$\,s \citep{patterson02}, looks remarkably similar to that of AM CVn \citep{roelofsamcvn}; it shows relatively simple `S-waves' in the \mbox{He\,{\sc i}} 4387, \mbox{He\,{\sc i}} 4471 and \mbox{He\,{\sc i}} 4921 lines, as well as in the \mbox{Mg\,{\sc ii}} 4481 line. Like in AM CVn and other interacting binaries, the strength of the S-wave is seen to vary strongly with phase, which can be explained by a varying visibility of the accretion stream--accretion disc impact region as is expected to occur quite naturally in an inclined system. No S-waves or other structure are seen when we fold the data on the suggested 1119-second superhump period; also in this respect HP Lib is identical to AM CVn \citep{nsg}. We thus conclude that our spectroscopy supports a 1102.7-second orbital period for HP Lib.

We again perform a Monte Carlo test to determine our bright spot phase resolution and the associated error on the orbital period that we determine. To this end we make bootstrapped Doppler tomograms from the spectra of night 1 and 2, and from the spectra of night 4 and 5. The baseline between these two data-subsets is 3 nights, and the relative bright spot phase resolution is measured to be about 10 degrees. Allowing again for a 10-degree intrinsic shift in the bright spot due to changes in the effective accretion disc radius, this gives a combined bright spot phase resolution of about 15 degrees, or a resolution in orbital period of 0.2 seconds. Best alignment of the bright spots in the data-subsets is achieved for a folding period of 1102.8 seconds. Our best measurement for the period is thus \ph.

\begin{figure}
\centering
\includegraphics[angle=270,width=84mm]{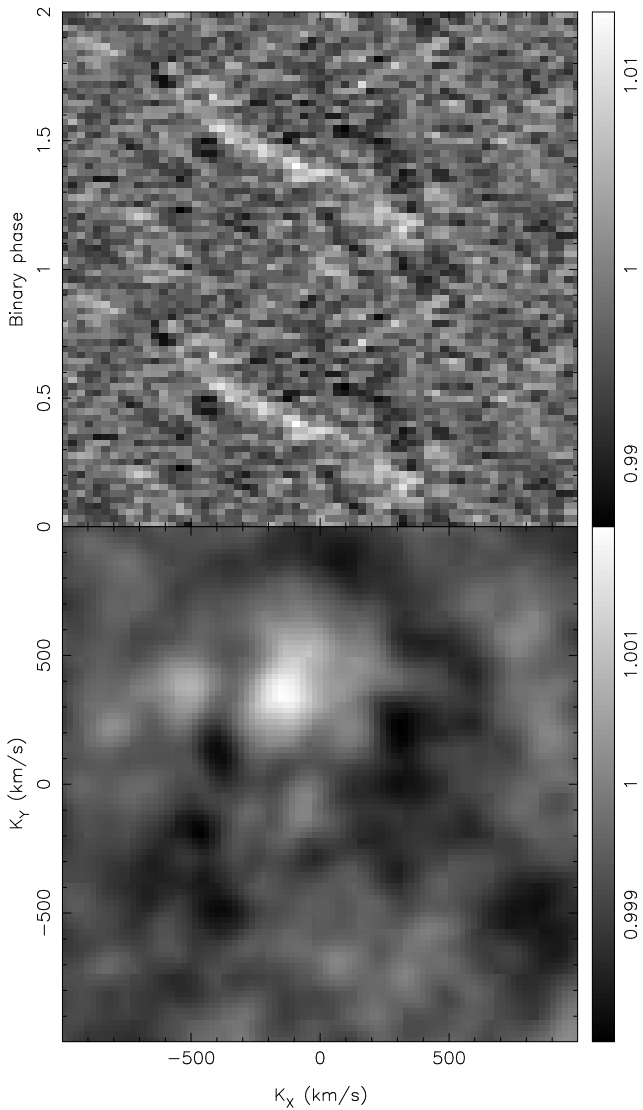}
\caption{Average-subtracted trailed spectrum (top) and corresponding linear back-projection Doppler tomogram of the combined lines \mbox{He\,{\sc i}} 4387, \mbox{He\,{\sc i}} 4471, \mbox{Mg\,{\sc ii}} 4481 and \mbox{He\,{\sc i}} 4921 in HP Lib, folded on $P_\mathrm{HP\,Lib}=1102.8$\,s. The grey-scale indicates the relative flux levels.}
\label{trail_doppmapHPLib}
\end{figure}

Since the S-wave features are rather weak, we show a combined spectrogram and Doppler tomogram of the four aforementioned spectral lines in figure \ref{trail_doppmapHPLib}. The velocity semi-amplitude of the bright spot and its uncertainty, again determined from our bootstrap tests, come out at $K_\mathrm{HP\,Lib}=360\pm20$ km/s.

\section{Discussion \& conclusions}

\subsection{The orbital periods of HP Lib and V803 Cen}

If we make the (crucial) assumption that the emission features we have identified in phase-resolved spectroscopy of HP Lib and V803 Cen are (almost) stationary in the binary frame, as is the case if they correspond to the `S-waves' or `bright spots' commonly associated with the impact of the accretion stream into the disc, we can identify our determined spectroscopic periods with the true orbital periods of these binaries. The orbital periods of HP Lib and V803 Cen then are \ph\ and \pv.

Our orbital period of HP Lib agrees perfectly with the value $1102.70\pm0.05$\,s found in the literature, which is derived from broad-band photometry \citep{patterson02}. Our orbital period of V803 Cen on the other hand is significantly shorter than the literature value of $1611\pm1$\,s, which is again derived from photometry (\citealt{patterson01,patterson02}; but see \citealt{patterson00} for its initial interpretation as a possible superhump). It is quite conceivable that the low inclination of V803 Cen \citep{donoghue89,nasser} causes the photometric modulation on the orbital period to be so small, that even in extensive data-sets only superhumps are observed. We note that the velocity semi-amplitude of the bright spot in our Doppler tomograms, $K_\mathrm{V803\,Cen}=170\pm15$\,km/s, agrees with a binary of low inclination if we assume that it corresponds to the classical bright spot caused by the accretion stream--accretion disc impact ($i\sim15^\circ$, see also \citealt{roelofshst}).

An interesting question would be whether our spectroscopic period of 1596.4 seconds for V803 Cen may correspond to a negative superhump, so that 1611 seconds can still be its orbital period and 1618 seconds the `normal' (positive) superhump. Although technically possible, we consider this unlikely. The spectroscopic S-wave behaviour observed in V803 Cen is seen in many interacting binaries and has, as far as we know, always been associated with the accretion stream--accretion disc impact, rather than with a negative superhump or another feature not rotating on the orbital period. In addition, negative superhumps have not been detected in photometric studies of V803 Cen, which further weakens their case even though it does not rule out their existence.

We thus believe that the 1611-second period, observed in the intermediate and low brightness states of V803 Cen, is probably a superhump, as concluded in the original paper by \citet{patterson00}. The same goes for the 1618-second period observed during V803 Cen's high brightness state \citep{patterson00}. Our spectroscopic S-wave period of \pv\ is most likely the underlying orbital period of the binary. It is unclear how exactly the 1611-second period has entered the literature as the orbital period of V803 Cen. As far as we know, it first appeared in data tables in \citet{patterson01} and \citet{patterson02}, but we were unable to find a publication in which the actual detection of an orbital signal (at 1611 seconds or any other period) is claimed.

\subsection{The superhump period excess and mass ratio}

It is well known empirically, as well as from numerical modelling, that there is a correlation between the superhump period excess $\epsilon$ and the mass ratio $q$ in Cataclysmic Variables \citep{patterson05,hirose,whitehurst88,whitehurst91}. Using our spectroscopic period of V803 Cen, \pv, as the orbital period, and combining with the superhump period of 1611--1618\,s measured by \citet{patterson00}, we arrive at a superhump period excess $\epsilon_\mathrm{V803\,Cen}=0.009$--$0.014$, comparable to the value $\epsilon_\mathrm{HP\,Lib}=0.015$ \citep{patterson02}. Combined with the most recent empirical relation between the superhump period excess and the mass ratio for the hydrogen-rich Cataclysmic Variables \citep{patterson05}, this would imply a mass ratio $q_\mathrm{V803\,Cen}=0.05$--$0.07$. This is already much larger than the value $q_\mathrm{V803\,Cen}=0.016$ found in the literature (e.g.\ \citealt{deloye}).

If we further allow for a different scaling between the superhump period excess and the mass ratio, as measured for AM CVn itself \citep{roelofsamcvn}, we obtain an even larger mass ratio $q_\mathrm{V803\,Cen}=0.08$--$0.11$. One could, of course, question the use of such $\epsilon(q)$ relations to derive mass ratios all too easily. The point remains, however, that there is little reason to assume that the mass ratio of V803 Cen is as extreme as previously thought, based on the orbital period that we propose here for V803 Cen.

\subsection{Constraints on the component masses}

\citet{deloye} derive very stringent limits on the masses of the stars in V803 Cen based on the small mass ratio $q_\mathrm{V803\,Cen}=0.016$. Basically, the minimum donor star mass is set by the mass of a cold, fully degenerate, Roche-lobe filling star, which works out to be $M_\mathrm{2,min}=0.021\,M_\odot$ for a helium donor in an orbit of about 1600 seconds. This then implies a massive accretor of minimum mass $M_\mathrm{1,min}=1.3\,M_\odot$. Conversely, the Chandrasekhar limit on the accretor mass implies a maximum donor star mass $M_\mathrm{2,max}=0.023\,M_\odot$; in other words, there is only a very narrow range of allowed masses for the donor star. From this it is inferred by \citet{deloye} that the donor star must be a cold, nearly perfectly degenerate white dwarf.

The mass ratio derived from our proposed orbital period and the associated superhump period excess relaxes these constraints almost completely, to $M_\mathrm{1,min}=0.3\,M_\odot$, $M_\mathrm{2,max}=0.10\,M_\odot$ for a mass ratio $q=0.07$. It would thus seem reasonable to assume that the donor star can be much heavier than a zero-temperature white dwarf. The large observed bolometric luminosity of V803 Cen can then relatively easily be explained by a hot, semi-degenerate donor star donating matter to a much less extreme accreting white dwarf of mass $M_1\lesssim1.0\,M_\odot$ (see \citealt{roelofshst}).

\section{Acknowledgments}

GHAR and PJG are supported by NWO VIDI grant 639.042.201 to P.J. Groot. DS acknowledges a Smithsonian Astrophysical Observatory Clay Fellowship. GN was supported by NWO VENI grant 639.041.405 to G. Nelemans. TRM was supported by a PPARC Senior Research Fellowship. This work is based on observations made at the European Southern Observatory, Chile, under programmes 072.D-0425(A) and 074.D-0662(A).

\end{document}